\documentclass[12pt,preprint]{aastex}
\def\arcsec{\hbox{$^{\prime\prime}$}}

\newcommand{\cmjj}{\mbox{${\rm cm^{-2}}$}}
\newcommand{\etal}{et al.}
\newcommand{\hI}{\mbox{${\rm H\ I}$}}
\newcommand{\kms}{\mbox{km\ s${^{-1}}$}}
\newcommand{\lya}{\mbox{${\rm Ly}\alpha$}}

\begin{document}

\def \cmm  {cm$^{-2}$}
\def \cmmm {cm$^{-3}$}
\def \kms  {km~s$^{-1}$}
\def \lyaf {Ly$\alpha$ forest}
\def \Lya  {Ly$\alpha$}
\def \lya  {Ly$\alpha$}
\def \Lyb  {Ly$\beta$}
\def \lyb  {Ly$\beta$}
\def \Lyg  {Ly$\gamma$}
\def \lyg  {Ly$\gamma$}
\def \ly5  {Ly-5}
\def \ly6  {Ly-6}
\def \ly7  {Ly-7}
\def \nhi  {$N_{HI}$}
\def \lnhi {$\log N_{HI}$}
\def \etal {\textit{et al.}}
\def \ob {$\Omega_b$}
\def \obh {$\Omega_bh^{-2}$}
\def \om {$\Omega_m$}
\def \ol {$\Omega_{\Lambda}$}
\def \gz {$g(z)$}
\def \lyaf {Lyman--$\alpha$ forest}
\def \hfreq {$f_{\rm{HI}} (N,X)$}

\title{A Shot in the Dark: A Technique for Locating the Stellar Counterparts of
Damped \lya\ Absorbers\altaffilmark{1}}

\author{
JOHN M.\ O'MEARA\altaffilmark{2}, HSIAO-WEN CHEN\altaffilmark{3}, and
DAVID L. KAPLAN\altaffilmark{2}
}
\altaffiltext{1}{Observations reported here were obtained at the
Magellan telescopes, a collaboration between the Observatories of the Carnegie
Institution of Washington, University of Arizona, Harvard University,
University of Michigan, and Massachusetts Institute of Technology.}
\altaffiltext{2}{MIT Kavli Institute for Astrophysics and Space Research, 
Massachusetts Institute of Technology, Cambridge, MA 02139, 
{\tt omeara@mit.edu, dlk@space.mit.edu}}
\altaffiltext{3}{Department of Astronomy \& Astrophysics, University of 
Chicago, Chicago, IL 60637, {\tt hchen@oddjob.uchicago.edu}}

\begin{abstract}

We present initial results from a Magellan telescope program to image
galaxies that give rise to damped \lya\ absorbers (DLAs) at  
$1.63 \le z_{\rm DLA} \le 2.37.$  
Our program differs from previous efforts in that we target 
quasars with intervening Lyman limit systems (LLS) 
along the line of sight at redshift $z_{\rm LLS} > 3.5$.  The
higher-redshift LLS is applied as a blocking filter to remove the
glare of the background quasar at the rest-frame ultraviolet
wavelengths of the foreground galaxy.  The complete
absence of quasar light offers an unimpeded view along the sightline
to the redshift of the LLS, allowing an exhaustive search for the DLA
galaxy to the sensitivity limit of the imaging data (at or better than
$0.25\,L_*$).  In both of our pilot fields (PKS2000$-$330, 
$z_{\rm DLA}=2.033$ and SDSS0322$-$0558, $z_{\rm DLA}=1.69$), we
identify an $L_*$ galaxy within $5\arcsec$ from the sightline which has
 optical colors consistent with star-forming galaxies at $z\approx
2$.  We examine the correlation between absorption-line properties and
galaxy luminosity and impact distance, and compare the high-redshift
galaxy and absorber pairs with those known at $z<1$.

\end{abstract}


\section{INTRODUCTION}

Damped \lya\ absorption systems (DLAs) identified along the line of
sight toward distant quasars 
offer an interesting alternative for finding
distant galaxies based on their neutral gas content, rather than
apparent brightness or color.  Despite a well-established chemical
enrichment history in the DLA population (e.g.\ Prochaska \etal\
2003), constraints for different star formation recipes from DLA
studies remain limited (e.g.\ Nagamine \etal\ 2006) because of a lack
of known stellar counterparts.  Specifically, the
low metal content observed in the DLA population is generally
interpreted as supporting evidence for an origin in the dwarf galaxy
population (e.g.\ Pettini \etal\ 1994), as opposed to an unbiased
sample of the field galaxy population (c.f.\ Chen \etal\ 2005).  At
the same time, studies of the fine structure transitions in the Si$^+$
and C$^+$ ions associated with DLAs (Wolfe \etal\ 2004) suggest that
DLAs may contribute equally to the star formation rate density at
redshift $z=3$ as the luminous starburst population selected at
rest-frame UV wavelengths (e.g.\ Steidel \etal\ 1999).

The nature of DLA galaxies can be determined directly from comparisons
between their luminosity distribution function with those of different
field populations.  
Identifying the absorbing galaxies allows us to
not only measure their intrinsic luminosity but also study the gaseous
extent of distant galaxies based on the galaxy-absorber pair sample.
At $z<1$, Chen \& Lanzetta (2003) showed that the luminosity
distribution function of 11 DLA galaxies is consistent with what is
expected from the general galaxy population, with a peak at $L \simeq
0.4 L_{*}$ and less than 40\% arising from $L \le 0.1 L_{*}$ galaxies.
This result clearly argues against a predominant origin of the DLAs in
dwarf galaxies.  At higher redshift, however, identifying DLA galaxies
becomes exceedingly difficult because of their still fainter
magnitudes and small angular separation from the background quasar.
To date, only six out of the $>500$ known DLAs have been uncovered in
stellar emission (see Table 1 of Weatherley et al., 2005 for a
summary).

We have initiated a multi-band imaging study of quasar fields with
known intervening absorbers at $z>1$ that exhibit strong Mg\,II,
Fe\,II, and sometimes Mg\,I absorption features and are therefore
promising DLA candidates ($> 60$\% likelihood; see Rao \etal\ 2006).
We have targeted our searches specifically in quasar fields that also
have an intervening Lyman limit system (LLS) along the line of sight
at $z_{\rm LLS} > z_{\rm DLA}$.  In each field, we employ the
higher-redshift LLS as a natural blocking filter of the background
quasar light at the rest-frame ultraviolet wavelengths of the
foreground DLA.  The complete absence of quasar light allows an
unimpeded search of intervening faint galaxies along the quasar
sightline to the redshift of the LLS, substantially increasing the
likelihood of finding the DLA galaxies.  The goal of our program is to
collect a large sample of DLA galaxies at $z>1$ for follow-up studies.
In this \textit{letter}, we present imaging results from two
pilot fields of our project.  Throughout the paper, we adopt a
$\Lambda$ cosmology, $\Omega_{\rm M}=0.3$ and $\Omega_\Lambda = 0.7$,
with a dimensionless Hubble constant $h = H_0/(100 \ {\rm km} \ {\rm
s}^{-1}\ {\rm Mpc}^{-1})$.

\section{PROGRAM DESIGN}

Our survey is designed to search for DLA galaxies in the absence of
background quasar light.  Therefore, we specifically target those DLA
fields for which an intervening Lyman limit system (LLS) also exists
along the line of sight.  The concept
is illustrated in Figure 1, where in the bottom panel we show the
spectrum a quasar at $z=3.778$.  An intervening LLS is present at
$z_{\rm LLS}=3.549$, which completely absorbs the photons from the
background quasar at wavelength $\lambda < 4150$ \AA.  A DLA is
identified at $z_{\rm DLA}=2.033$ based on the presence of strong
Mg\,II $\lambda\,2796,2803$ doublet and Mg\,I 2852 at $\lambda > 8400$
\AA\ (Figure 2).

Together with the throughput curves of typical broad-band filters, the
bottom panel of Figure 1 clearly indicates that the intervening LLS
serves as an additional blocking filter of the background quasar light
in the $u'$ band.  This particular DLA-LLS-QSO combination along a
single sightline offers a unique opportunity for conducting an
exhaustive search for the DLA galaxy in the $u'$ band without the
interference of background quasar light.  For an LLS at sufficiently
high redshift (i.e.\ $z_{\rm LLS} \ge 3.5$), this design allows us to
search for ultraviolet emission from DLA galaxies at $1.63 \le z_{\rm
DLA} \le 2.37$ with $\le 0.001$\% contaminating quasar flux in the
$u'$ band.  We caution that for the technique to be successful, either
the $u'$ filter must suppress red leak to approximately one part in $10^4$ or
a red blocking filter must also be used.

Our absorber sample is selected based on the presence of strong metal
absorption features that are indicative of DLA, because the presence
of the intervening LLS prevents us from measuring the neutral hydrogen
column density $N(\hI)$.  Specifically, a combination of strong
absorption in both Mg\,II and Fe\,II with rest-frame absorption
equivalent width $W_r > 0.5$ \AA\ (Figure 2) has also been employed by
Rao \& Turnshek (2000) and further refined in Rao \etal\ (2005) for
selecting DLAs.  We note that these strong metal-lines guarantee that the absorbers originate in
either a canonical DLA with $N(\hI) \ge 2 \times 10^{20}$ \cmjj\ and Fe abundance $\left[ \rm{Fe} / \rm{H} \right] = -1.0$ (see
e.g.\ Prochaska \etal\ 2003) or in a region with lower $N(\hI)$ but
near or greater than solar metallicity. 
In both scenarios, these strong metal-line absorbers offer a means of
identifying star-forming galaxies at high redshift based on the metal
content in their ISM, rather than their optical brightness or color.

\section{OBSERVATIONS}

We have completed a multi-band imaging program for two pilot fields
toward PKS2000$-$330 ($z_{\rm QSO}=3.778$) and SDSS0322$-$0558
($z_{\rm QSO}=3.945$).  Properties of the intervening absorbers along
these sightlines are summarized in columns (3) through (6) of Table 1.
Absorption-line profiles of metal transitions identified with the
candidate DLAs are presented in Figure 2.  The spectrum of
PKS2000$-$330 was obtained using the MIKE echelle spectrograph
(Bernstein \etal\ 2003) on the 6.5 m Magellan Clay telescope at Las
Campanas Observatory.  
In addition to the prominent LLS found at
$z_{\rm LLS}=3.550$ (Figure 1), a candidate DLA is found at $z_{\rm
DLA}=2.033$ based on the presence of Mg\,I, Mg\,II and Fe\,II.
The presence of Mg\,I indicates that the metal-line
transitions trace neutral gas and therefore provides further support
for a DLA nature of this absorber.
The spectrum of SDSS0322$-$0558 was retrieved from the SDSS data
archive. 
In addition to an LLS found at $z_{\rm LLS}=3.764$ in the SDSS
quasar spectrum, a candidate DLA is found at $z = 1.691$ based on
detections of both strong Mg\,II and Fe\,II.  Because of the low
resolution of SDSS coupled with poor signal to noise in this
particular spectrum, we were unable to observe other metal line
transitions.

Optical imaging observations of the two quasar fields were carried out
using the MagIC direct imager on the Magellan Clay telescope with the
$u'$, $g'$, and $r'$ filters.  The optical imaging data presented in
this paper were obtained separately in three different runs in August,
September, and October of 2005. 
All these images were taken under photometric conditions with an
exquisite mean seeing of 0.6\arcsec, 0.5\arcsec, and 0.7\arcsec\ in
the $u'$, $g'$, and $r'$ bands, respectively.  Object photometry was
calibrated using Landolt standards (Landolt 1992) observed on the same
nights as the science frames.

Optical $u'$, $g'$, and $r'$ images of the field around
SDSS0322$-$0558 were obtained in September 2005
through cirrus, and object photometry was calibrated using
common sources identified in the SDSS field.  The mean seeing in 
the final stacked $u'$, $g'$, and $r'$ images are respectively 
0.8\arcsec, 0.7\arcsec, and 0.7\arcsec.  

Near-infrared images of the field around SDSS0322$-$0558 were also
obtained in October 2005, using Persson's Auxiliary Nasmyth Infrared
Camera (PANIC; Martini \etal\ 2004) on the Magellan Baade telescope
with the $H$ filter.  
The images were taken under photometric conditions with a mean
seeing of 0.4\arcsec.
Object photometry was calibrated using several
Persson infrared standards (Persson \etal\ 1998) observed on the
same nights of the science frames.

All imaging data were processed using standard pipeline techniques.
The processed individual images were registered to a common origin,
filtered for deviant pixels, and stacked to formed a final combined
image using our own program.  For each field, we detect objects
separately in the $u'$ and $g'$ frames due to the absence and presence
of the background quasar in these two bandpasses, using the
SExtractor program (Bertin \& Arnout 1996).
A segmentation map is produced that defines the sizes and shapes of all
the objects found in the stacked images.  Object fluxes were measured
by summing up all the photons in the corresponding apertures in the
segmentation map.  Flux uncertainties were estimated from the mean
variance over the neighboring sky pixels.  The dominant error in
object photometry is, however, in the photometric zero-point calibration.
We estimate uncertainties in optical and near-infrared photometry are
0.1 mag and 0.05 mag, respectively.

\section{Results and Discussion}

The results for PKS2000$-$330 and
SDSS0322$-$0558 are presented in Figures 1 and 3.  Galaxies identified
at angular distance $\theta < 5\arcsec$ are summarized in columns (7)
through (12) of Table 1.  

In the field of PKS2000$-$330, two extended sources, $X$ and $G$, are
found close to the sightline and above the $5\,\sigma$ limiting
magnitude $AB(u')=26.9$ over a 1\arcsec-diameter aperture.  Both $X$
and $G$ have similar $u'$-band brightness $AB(u')=25.4\pm 0.1$ and
$u'-g'$ color, $AB(u'-g')=0.4$.  Object $X$ is located at 2.8\arcsec\
from the sightline, not resolved from the quasar in the $r'$-band
image, and object $G$ is located at 4.7\arcsec\ from the sightline.
We note that galaxies $X$ and $G$ are not likely associated with the
strong LLS at $z_{\rm LLS} = 3.2$ that produces the strong \lya\
absorption feature at $\lambda\approx 5100$ \AA, because at $z=3.2$
nearly all the photons at $\lambda < 3800$ (rest-frame 912 \AA) are
absorbed and the abundant fluxes observed in the $u'$-band would imply
an enormous ultraviolet flux intrinsic to these galaxies.  Figure 4
shows that the observed colors of galaxy $G$ are consistent with
star-forming galaxies at $1.9<z<2.7$ (Adelberger \etal\ 2004).  We
therefore consider galaxy $G$ (and possibly $X$) as the DLA galaxy at
$z_{\rm DLA}=2.033$ and the $r'$-band magnitude of $G$ indicates that
it is nearly an $L_*$ galaxy (c.f.\ $AB_*(R)=24.54$ for $z=3$ galaxies
in Adelberger \& Steidel 2000).  At $z_{\rm DLA}=2.033$, the
corresponding impact parameter of $X$ and $G$ is $\rho=16.3\ h^{-1}$
kpc and $27.5\ h^{-1}$ kpc, respectively.

In the field of SDSS0322$-$0558, we identify a single compact source
$G$ at $\theta=2.1\arcsec$ to the quasar with $AB(u')=23.8\pm 0.1$,
and no other sources brighter than the $5\,\sigma$ limiting magnitude
$AB=26.8$ in the $u'$-band.  In Figure 4, we show that this object has
observed colors consistent with star-forming galaxies at $1.9<z<2.7$.
We therefore consider this object the most likely candidate for the
DLA at $z_{\rm DLA}=1.69$.  At $z=1.69$, the galaxy is at $\rho=12.4\
h^{-1}$ kpc and its $H$-band magnitude also indicates a nearly $L_*$
galaxy (e.g.\ Chen \etal\ 2003).

We have shown that in the absence of quasar light the $u'$-band image
of each field offers an unimpeded view along the line of sight to the
redshift of the LLS, allowing a complete search of the DLA galaxy to
the sensitivity limit of the imaging data.  The consistent optical
colors of these candidate galaxies summarized in Table 1 strongly
support their identification as the absorbing galaxies, despite a lack
of redshift measurements.  We note that the sensitivity of the
$u'$-band images allows us to uncover galaxies fainter than $L_*$
($>0.25\,L*$ in PKS2000$-$330 and $>0.06\,L_*$ in SDSS0322$-$0558) at
the redshift of the DLA, but no fainter galaxies are found.
Our identification is further strengthened by the number of random
galaxies we expect to find within the small angular radius from the
quasar along the quasar sightline.  Adopting a nominal galaxy
luminosity function from Ellis \etal\ (1996), we expect to find $<
0.1$ galaxies with $AB(u')<=25.4$ within $\theta\le\,5\arcsec$ over
the redshift interval $\Delta\,z=0-z_{\rm LLS}$.

Our survey results based on two Mg\,II-selected DLAs at $z>1.5$ agree
with previous studies at $z<1$ in that strong Mg\,II absorbers are
commonly found to arise in typical $L_*$ galaxies (Bergeron 1986;
Steidel \etal\ 1994).  Their intrinsic luminosities are also
comparable to the few DLA galaxies known at $z>1.9$ (M{\o}ller \etal\
2002).  This is, however, at odds with the null results reported by
Colbert \& Malkan (2002) and with recent work by Rao \etal\ (2003),
who argued that strong Mg\,II-selected DLAs at $z<1$ most likely
originate in $< 0.1\,L_*$ dwarf galaxies.  Using the two
galaxy-absorber pairs established in our pilot study, we also find
that the $W_r({\rm Mg\,II~2976})$ versus $\rho$ distribution at
$z>1.6$ is qualitatively consistent with the anti-correlation observed
at $z<1$ (Lanzetta \& Bowen 1990; Churchill \etal\ 2000).  The
agreement suggests that the size of extended Mg\,II gas around
luminous galaxies has not evolved significantly since $z=1.6$.  A
larger sample is necessary for investigating the extent of neutral gas
in high-redshift galaxies, as well as for a statistical comparison
between the luminosity distribution of the DLA galaxies and that of
the field galaxy population.

\acknowledgments

  We appreciate the expert assistance from the staff of the Las
Campanas Observatory.  We thank Rob Simcoe and Paul Schechter for obtaining some of the imaging data for the project.   It is a pleasure to thank
Rob Simcoe and Scott Burles for their helpful comments on earlier drafts of
this paper. J.M.O. acknowledges partial support from NSF grant AST-0307705.
H.-W.C. acknowledges partial support from NASA grant NNG06GC36G.

\clearpage

\begin{deluxetable}{p{0.85in}cccccccccccc}
\rotate
\tablewidth{0pc} \tablecaption{Summary of Known Properties of
Candidate DLAs} \tabletypesize{\scriptsize} \tablehead{ \colhead{} &
\colhead{} & \multicolumn{4}{c}{Absorbers} & \colhead{} &
\multicolumn{6}{c}{Galaxies\tablenotemark{b}} \\ 
\cline{3-6}
\cline{8-13} \\ 
\colhead{} & \colhead{} & \colhead{} & \colhead{} &
$W_{r}(\AA)$ & $W_{r}(\AA)$ & \colhead{} & \colhead{$\theta$} &
\colhead{$\rho$} & \colhead{} & \colhead{} & \colhead{} & \colhead{} \\ 
\colhead{Field} & \colhead{$z_{\rm QSO}$} & \colhead{$z_{\rm LLS}$} &
\colhead{$z_{\rm DLA}$} & (${\rm Mg\,II}_{2796}$) & (${\rm
Fe\,II}_{2600}$) & \colhead{} & \colhead{(\arcsec)} &
\colhead{($h^{-1}\,{\rm kpc}$)} & \colhead{$AB(u')$} &
\colhead{$AB(g')$} & \colhead{$AB(r')$} & \colhead{$AB(H)$} \\
\colhead{(1)} & \colhead{(2)} & \colhead{(3)} & \colhead{(4)} &
\colhead{(5)} & \colhead{(6)} & \colhead{} & \colhead{(7)} &
\colhead{(8)} & \colhead{(9)} & \colhead{(10)} & \colhead{(11)} &
\colhead{(12)} } 
\startdata 
\multicolumn{1}{r}{PKS2000$-$330($X$)} & 3.778 & 3.550 & 2.033 & 
$0.958 \pm 0.008$ & $0.513 \pm 0.006 $ & & 2.8 & 16.4 & $25.4$ & 
$25.0$ & ... & ... \nl 
\multicolumn{1}{r}{($G$)} & ... & ... & ... & ... & ... & & 
4.7 & 27.5 & $25.4$ & $25.0$ & $24.9$ & ... \nl 
SDSS0322$-$0558 & 3.945 & 3.764 & 1.690 & $4.3 \pm 0.3\tablenotemark{b}$ 
& $1.7 \pm 0.2$ & & 2.1 & 12.4 & $23.8$ & $23.3$ & $23.0$ & $22.08$ 
\enddata 
\tablenotetext{a}{The error on this measurement is possibly higher, 
due to a noise spike in the absorption feature.}
\tablenotetext{b}{Uncertainties in galaxy magnitude are dominated by
systematic uncertainties in the photometric zero-point calibration.
We estimate uncertainties in optical and near-infrared photometry are
0.1 mag and 0.05 mag, respectively.}
\end{deluxetable}

\clearpage

\begin{figure*}[th]
\begin{center}
\includegraphics[scale=0.6]{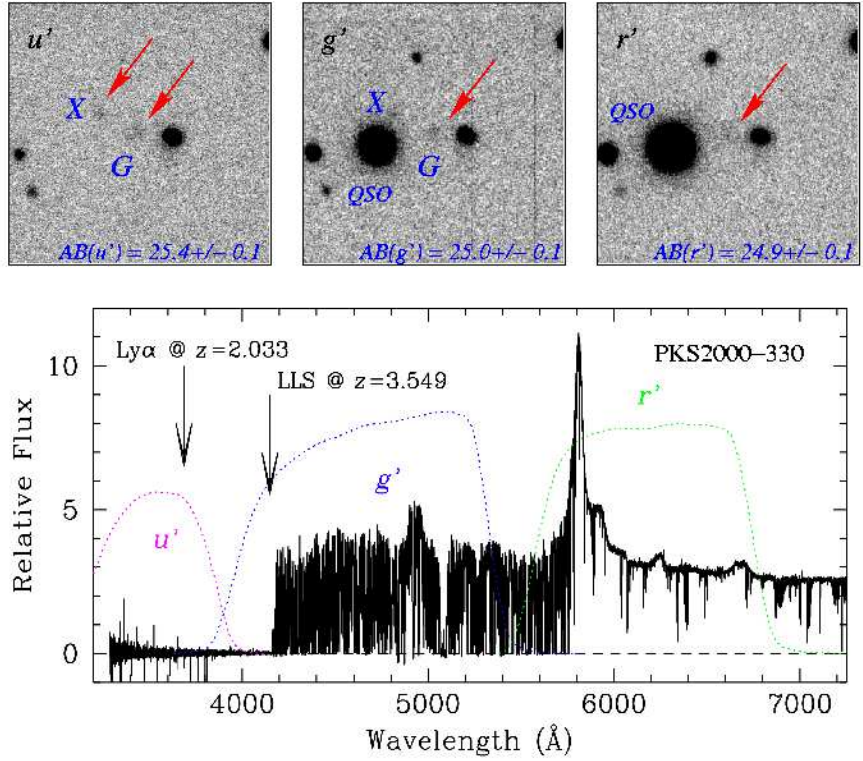}
\caption{An example of one of our survey fields toward PKS2000$-$330
($z_{\rm QSO}=3.778$).  The bottom panel illustrates the design of our
survey.  A high-resolution echelle spectrum of the quasar is displayed
in solid, black curve, where the redshifted \lya\ emission feature is
present at 5800 \AA.  The arrows indicate along the sightline the
location of a LLS at $z_{\rm LLS}=3.549$ and the expected \lya\
feature for the DLA candidate at $z_{\rm DLA}=2.033$, as selected by
the presence of strong Mg\,II and Fe\,II at $\lambda > 8000$ \AA\ (see
Figure 2).  Superimposed are bandpasses for typical $u'$, $g'$, and
$r'$ filters (dotted curves).  The top panels show deep images of the
quasar field in the corresponding $u'$, $g'$, and $r'$ bands, where
the quasar light is completely missing in the $u'$ band due to the
presence of the intervening LLS.  The images are 20\arcsec\ on a side.
We have identified two galaxies $X$ and $G$ at 2.8\arcsec\ and
4.7\arcsec, respectively, from the quasar sightline.  We present the
broad-band magnitude of object $G$ in the lower-right corner of each
image panel.}
\end{center}
\end{figure*}

\begin{figure}[th]
\begin{center}
\includegraphics[scale=0.45]{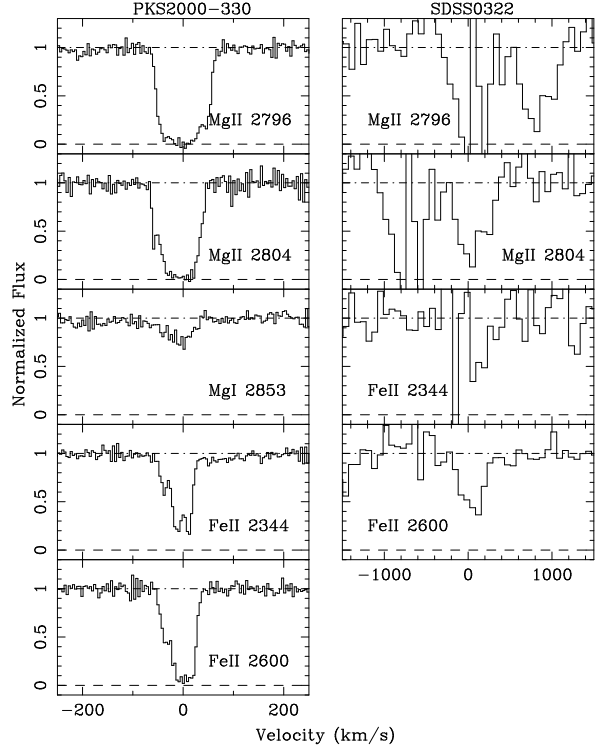}
\caption{Absorption-line profiles of various heavy elements identified
along the sightlines toward PKS2000$-$330 at $z_{\rm QSO}=3.778$ (left
panels) and SDSS0322$-$0558 at $z_{\rm QSO}=3.945$ (right panels).
The zero relative velocity corresponds to $z=2.033$ for the absorber
toward PKS2000$-$330 and $z=1.691$ for the absorber toward
PKS0322$-$0558.}
\end{center}
\end{figure}

\begin{figure}[th]
\vskip 0.1in
\begin{center}
\includegraphics[scale=0.4]{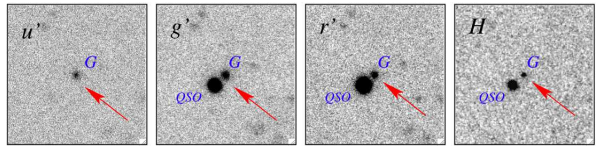}
\caption{Stacked images of field around SDSS0322$-$0558 ($z_{\rm
QSO}=3.945$) in the $u'$, $g'$, and $r'$ bands, where the quasar light
is again completely missing in the $u'$ band due to the presence of an
intervening LLS at $z_{\rm LLS}=3.764$.  The images are 20\arcsec\ on
a side.  We have identified a galaxy $G$ at 2.1\arcsec\ angular
distance from the quasar sightline.}
\end{center}
\end{figure}

\begin{figure}[th]
\begin{center}
\includegraphics[scale=0.4]{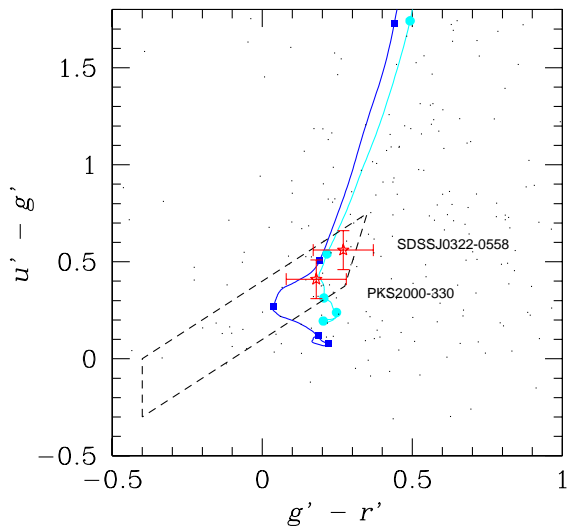}
\caption{Optical $u'-g'$ vs. $g'-r'$ colors of galaxies at $z\approx 2$. 
Observations of the candidate DLA galaxy $G$ in each of the survey fields
toward PKS2000$-$330 and SDSS0322$-$0558 are shown in open stars with 
error bars, together with random galaxies found in the two fields (dots).
The color criteria for selecting $1.9<z<2.7$ galaxies from Adelberger 
\etal\ (2004) are marked in dashed lines.  Solid curves are predicted
optical colors for starburst galaxies at high redshift under a no-evolution
scenario, starting at $z=1$ through $z=3$ in steps of $\Delta\,z=0.5$.
Both candidate DLA galaxies have observed colors consistent with the
selection criteria for $z\approx 2$ star-forming galaxies.}
\end{center}
\end{figure}

\end{document}